\documentclass[aps,twocolumn,superscriptaddress,amsmath,amssymb,nofootinbib,nobibnotes,prl,floatfix]{revtex4-2}
\pdfoutput=1 
\usepackage[utf8]{inputenc}
\usepackage[T1]{fontenc}
\usepackage[version=3]{mhchem}
\usepackage{graphicx}  % needed for figures
\usepackage{dcolumn}   % needed for some tables
\usepackage{bm}        % for math
\usepackage[english]{babel}
\usepackage[dvipsnames]{xcolor}
\usepackage[
colorlinks=true,
urlcolor=RoyalBlue,
linkcolor=RoyalBlue,
citecolor=RoyalBlue,
]{hyperref}
\usepackage{vmargin}
\setpapersize{A4} \oddsidemargin20mm \textwidth170mm \topmargin10mm
\usepackage[type1]{libertine}% Linux Libertine für zweispaltige Texte
\usepackage{textcomp}% Required to get special symbols
\usepackage[scaled=.85]{beramono}% Typewriter font
\usepackage[libertine,cmintegrals,cmbraces,vvarbb]{newtxmath} %,slantedGreek
\usepackage[scr=boondoxo]{mathalfa}% Extra math symbols
\usepackage[lf]{carlito}

\begin{document}
\title{Scanning Tunnelling Microscopy for Molecules: Effects of Electron\\ Propagation into Vacuum}
\date{\today}
\author{Abhishek Grewal}
%\email{a.grewal@fkf.mpg.de}
\author{Christopher C. Leon}
%\email{ccleon@mit.edu}
\author{Klaus Kuhnke}
\email{k.kuhnke@fkf.mpg.de}
\affiliation{Max-Planck-Institut f\"ur Festk\"orperforschung, Heisenbergstra{\ss}e 1, 70569 Stuttgart, Germany}
\author{Klaus Kern}
\affiliation{Max-Planck-Institut f\"ur Festk\"orperforschung, Heisenbergstra{\ss}e 1, 70569 Stuttgart, Germany}
\affiliation{Institut de Physique, {\'E}cole Polytechnique F{\'e}d{\'e}rale de Lausanne, 1015 Lausanne, Switzerland}
\author{Olle Gunnarsson}
\email{o.gunnarsson@fkf.mpg.de}
\affiliation{Max-Planck-Institut f\"ur Festk\"orperforschung, Heisenbergstra{\ss}e 1, 70569 Stuttgart, Germany}

%###################################################################%
%##################          ABSTRACT       ########################%
%###################################################################%

\begin{abstract}
\noindent Using scanning tunneling microscopy (STM), we experimentally and theoretically investigate isolated platinum phthalocyanine (PtPc) molecules adsorbed on atomically thin NaCl(100) vapor deposited on Au(111).
We obtain good agreement between theory and constant-height STM topography.
We examine why strong distortions of STM images occur as a function of distance between molecule and STM tip.
The images of the highest occupied molecular orbital (HOMO) and the lowest unoccupied molecular orbital (LUMO) exhibit, for increasing distance, significant radial expansion due to electron propagation in the vacuum.
Additionally, the imaged angular dependence is substantially distorted.
The LUMO image has substantial intensity along the molecular diagonals where PtPc has no atoms.
In the electronic transport gap the image differs drastically from HOMO and LUMO, even at energies very close to these orbitals.
As the tunneling becomes increasingly off-resonant, the eight angular lobes of the HOMO or of the degenerate LUMOs diminish and reveal four lobes with maxima along the molecular axes, where both, HOMO and LUMO have little or no weight.
These images are strongly influenced by low-lying PtPc orbitals that have simple angular structures.
\end{abstract}

\date{December 7, 2023}

\maketitle
%###################################################################%
%################          INTRODUCTION       ######################%
%###################################################################%
\clearpage
\section*{Introduction}\label{sec:1}
\noindent Scanning tunneling microscopy (STM) was developed across several papers by Binnig \textit{et al.}
\cite{binnigTunnelingControllable1982Appl.Phys.Lett., binnigSurfaceStudies1982Phys.Rev.Lett., binnig111Facets1983SurfaceScienceLetters, binnigIfmmodeTimes1983Phys.Rev.Lett.}
The technique has been extensively discussed in reviews \cite{binnigScanningTunneling1987Rev.Mod.Phys., drakovaTheoreticalModelling2001Rep.Prog.Phys., hoferTheoriesScanning2003Rev.Mod.Phys., eversAdvancesChallenges2020Rev.Mod.Phys., chenIntroductionScanning2021a}, as well as in early theoretical works \cite{tersoffTheoryApplication1983Phys.Rev.Lett., tersoffTheoryScanning1985Phys.Rev.B, garciaModelTheory1983Phys.Rev.Lett., baratoffTheoryScanning1984PhysicaB+C, stollResolutionScanning1984SurfaceScience, langVacuumTunneling1985Phys.Rev.Lett., langTheorySingleAtom1986Phys.Rev.Lett., chungSphericalTip1987SurfaceScience, sacksSurfaceTopography1987Phys.Rev.B, sacksTunnelingCurrent1988Phys.Rev.B, chenTheoryScanning1988J.Vac.Sci.Technol.A, chenOriginAtomic1990Phys.Rev.Lett., chenTunnelingMatrix1990Phys.Rev.B, sacksGeneralizedExpression1991Phys.Rev.B, sacksTersoffHamann1991J.Vac.Sci.Technol.BMicroelectron.NanometerStruct.Process.Meas.Phenom., sestovicContributionsHigher1995Phys.Rev.B}.
Significant theoretical progress was achieved by Tersoff and Hamann \cite{tersoffTheoryApplication1983Phys.Rev.Lett., tersoffTheoryScanning1985Phys.Rev.B}.
They assumed that the electrons tunnel to or from an $s$-orbital on the tip.
Using the Bardeen theory \cite{bardeenTunnelingSuperconductors1962Phys.Rev.Lett.}, they demonstrated that the tunneling current is determined by the hypothetical value of the wave function for the tunneling electron at the center of the $s$-orbital.
Substantial effort has been dedicated to improving this assumption \cite{baratoffTheoryScanning1984PhysicaB+C, chungSphericalTip1987SurfaceScience, sacksSurfaceTopography1987Phys.Rev.B, sacksTunnelingCurrent1988Phys.Rev.B, chenTheoryScanning1988J.Vac.Sci.Technol.A, chenOriginAtomic1990Phys.Rev.Lett., chenTunnelingMatrix1990Phys.Rev.B, sacksGeneralizedExpression1991Phys.Rev.B, sacksTersoffHamann1991J.Vac.Sci.Technol.BMicroelectron.NanometerStruct.Process.Meas.Phenom., sestovicContributionsHigher1995Phys.Rev.B, bodeDistancedependentSTMstudy1996Z.Phys.B-CondensedMatter, gottliebBardeenTunnelling2006Nanotechnology, chaikaSelectingTip2010EPL, palotasOrbitaldependentElectron2012Phys.Rev.B, grushkoAtomicallyResolved2013Nanotechnology, mandiChenDerivative2015Phys.Rev.B}.
However, these refinements necessitate a thorough understanding of the electronic structure of the tip.
Given the limited information available, we adopt the assumption of Tersoff and Hamann. 
Repp \textit{et al.} \cite{Repp} introduced a NaCl buffer when studying a molecule to reduce the influence of the substrate. 
They observed that STM shows spatially expanded images of molecular orbitals (MO) of a pentacene molecule \cite{Repp}.

\textit{Ab initio} calculations of STM images have been performed for clean surfaces and for very small molecules adsorbed directly on metal surfaces \cite{langVacuumTunneling1985Phys.Rev.Lett., langTheorySingleAtom1986Phys.Rev.Lett., pazEfficientReliable2006Phys.StatusSolidiB, teobaldiIncludingProbe2007Phys.Rev.B, rossenLowestOrder2013Phys.Rev.B, zhangEfficientMethod2014J.Phys.Chem.A, gustafssonTheoreticalModeling2017, gustafssonScanningTunneling2016Phys.Rev.B, okabayashiInfluenceAtomic2016Phys.Rev.B, gustafssonAnalysisSTM2017Phys.Rev.B, gustafssonSTMContrast2017J.Phys.:Condens.Matter}.
More closely related to the present work are studies of single copper  phthalocyanine (CuPc) molecules
using a model that includes the highest occupied MO (HOMO), the lowest unoccupied MO (LUMO) and one $\sigma$-orbital, located just below the energy of the CuPc HOMO \cite{siegertEffectsSpin2015BeilsteinJ.Nanotechnol., donariniTopographicalFingerprints2012Phys.Rev.B, siegertNonequilibriumSpin2016Phys.Rev.B}.
CuPc has been widely studied due to its demonstration of negative differential resistance \cite{tuControllingSingleMolecule2008Phys.Rev.Lett.}, while the related compound \ce{H2Pc} has garnered significant interest due to observations of up-conversion electroluminescence at tunneling energies in the \ce{H2Pc} transport gap \cite{chenSpinTripletMediatedUpConversion2019Phys.Rev.Lett., farrukhBiaspolarityDependent2021, GrewalSingleMolecule2022}.
These experiments raise important and fundamental questions about electrons tunneling through the transport gap.

\begin{figure}
    \includegraphics [width=200pt]{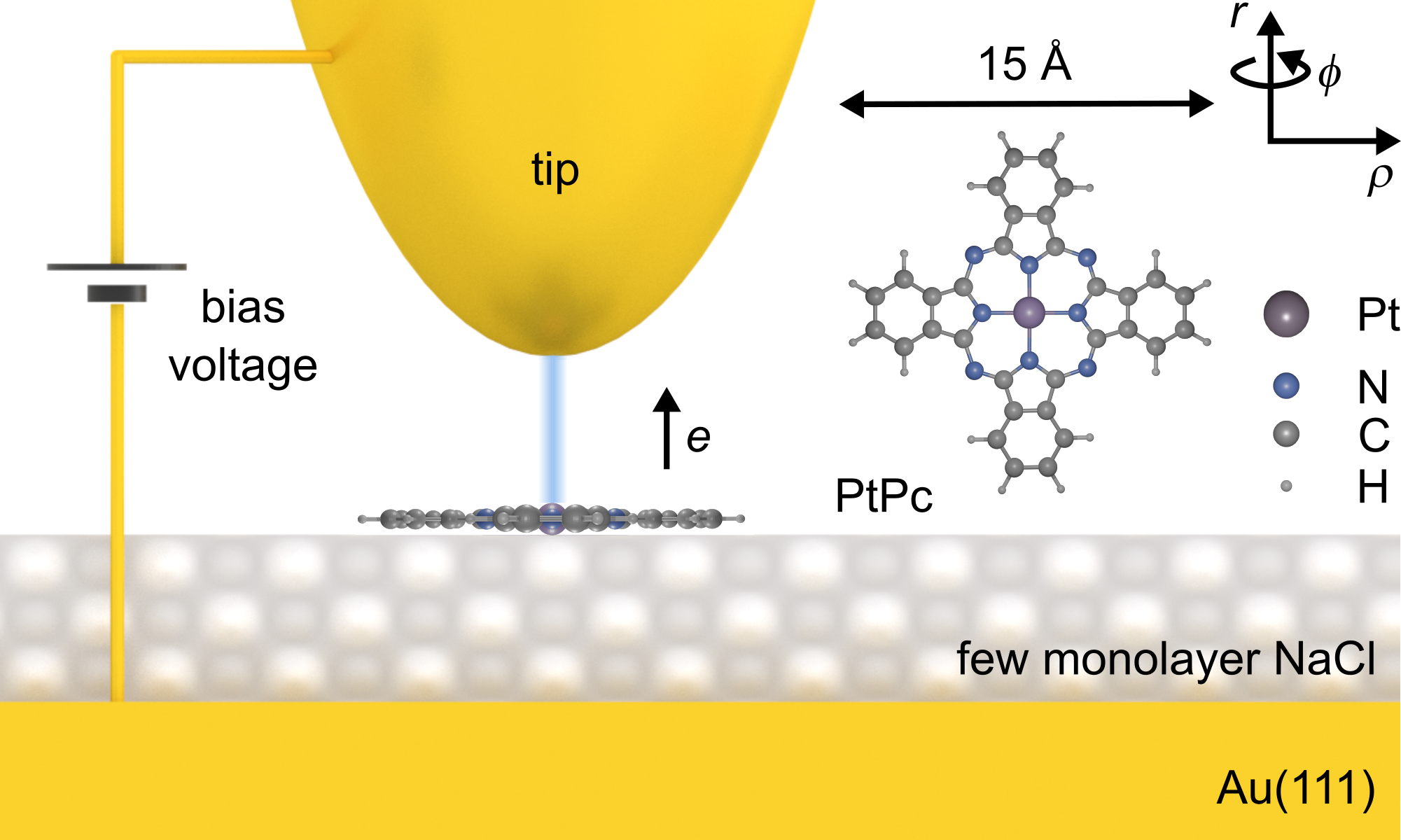}
	\caption{Scheme of the system investigated experimentally and simulated theoretically: PtPc on a thin NaCl(100) film on Au(111) in the STM tunnel junction. The four ``arms'' of the molecule are the isoindole units, shown aligned with the cardinal directions.}
	\label{fig:1}
\end{figure}

Here, we study a similar organic luminescent system, namely PtPc adsorbed on a few layers of a NaCl(100) 
film on an Au substrate, as illustrated in Fig. \ref{fig:1}.
In particular our study serves as a means to better understand the very nature of tunneling through vacuum in STM studies.
We show how images of the HOMO and the LUMO substantially expand radially and explain their origin.
The LUMO image also distorts angularly and has significant weight even at spatial positions where the underlying molecule has no atoms, for reasons to be discussed below.
The images of electrons tunneling through the transport gap are shown to differ drastically from the HOMO or LUMO images, even for energies very close to these states.
We show that in an expansion in terms of MOs this is due to contributions from MOs with few nodal surfaces at substantially lower energy than the HOMO.
This effect is very important for understanding experiments manipulating electrons (tunneling through the gap) and photons. 

Here we treat a single PtPc molecule adsorbed flat on the surface.
We expand the wave function in vacuum in cylindrical coordinates.
In planes parallel to the surface we use trigonometric functions, $\cos(m\phi)$ and $\sin(m\phi)$, to describe the angular behavior and integer Bessel functions, $J_m(k_{mi}\rho)$ with $i-1$ radial nodes in the region studied, to describe the radial behavior.
Perpendicular to the surface the wave function is described by exponentially decaying functions.
For the Au-NaCl-PtPc system we employ a tight-binding (TB) formalism \cite{leonAnionicCharacter2022NatCommun, grewalGapState2023ACSNano}.

We investigate electrons arriving at the tip with a given total kinetic energy, comprising positive 
contributions from the angular and radial variations parallel to the surface and negative contributions 
from the exponentially decaying part perpendicular to the surface.
We study these effects for the HOMO, the LUMO, and states in the transport gap. 

Basis states with a) radial functions with many nodes (large $k_{mi}$), or b) angular functions with many nodes (large $m$) have large (positive) kinetic energies in planes parallel to the surface.
These basis states are combined with exponential functions perpendicular to the surface, which have large negative kinetic energies, to obtain states with the total kinetic energy corresponding to the bias. 
These states decay rapidly in the perpendicular direction and play a small role for the topography at the tip.
This effect exponentially favors components with a) small $k_{mi}$ and b) small $m$, i.e., components with few nodal surfaces in the angular and radial parts of the wave function. 

Factor a) results in a significant expansion of the STM image in the radial direction when the smallest significant value of $m$ is not too small.
For PtPc, this phenomenon is applicable to both the HOMO and LUMO.
To observe the effects of factor b), we notice that PtPc has four ``arms'' along the $x$- and $y$-axis.
The HOMO exhibits a total of eight lobes, with a pair of lobes tending to align along each of the molecular arms.
Due to factor b) this tendency diminishes during electron propagation into vacuum.
The two-fold degenerate LUMOs collectively form a set of eight lobes.
Due to factor b) its image markedly accumulates intensity between the molecular arms in the $x\pm y$-directions, even though there are no atoms in the underlying molecule in these directions.
While factor b) tends to position the eight lobes of the HOMO at equal angles, the LUMO has a tendency to overshoot.      

Propagation through the transport gap becomes important for electron and photon manipulations.
Factors a) and b) mentioned above exponentially favor contributions from basis functions with few angular and radial nodes.
In Ref.~\citenum{grewalGapState2023ACSNano} we emphasized how propagation through the Au-NaCl-PtPc system 
also favors this type of basis states. Consequently, even for energies very close to the HOMO or LUMO, the in-gap 
image differs qualitatively from the HOMO and LUMO images. While the HOMO, as well as the two degenerate LUMOs 
together, have eight lobes, the in-gap image has only four lobes. Despite the HOMO having no intensity right on 
the $x$- or $y$-axes, and the LUMO having little intensity as well, the image in the gap has its four lobes 
centered on these axes. Specifically, in an expansion in terms of MOs, we find that the lowest $\pi$ orbital, 
with no nodal planes perpendicular to the surface, has a large weight due to its small kinetic energy parallel 
to the surface.  More detailed information about the model can be found in the Supplementary Information (SI). 
The next section provides a detailed description of the theoretical formalism. In the following section, we 
present the experimental and theoretical results.

\subsection{Theoretical formalism}\label{sec:3}
We study a PtPc molecule adsorbed on a NaCl(100) film of three atomic layers on a Au(111) substrate.
As in earlier work \cite{leonAnionicCharacter2022NatCommun, grewalGapState2023ACSNano}, we use a TB formalism to describe the Au-NaCl-PtPc system, essentially following prescriptions of Harrison \cite{harrisonElementaryElectronic1999} (see SI).
The resulting Hamiltonian is
\begin{eqnarray}\label{eq:1}
	&&H=\sum_{i\sigma} \varepsilon_i^{\rm Au}n_{i\sigma}^{\rm Au} 
	+\sum_{i\sigma} \varepsilon_i^{\rm NaCl}n_{i\sigma}^{\rm NaCl} 
	+\sum_{i=1}^{182}\sum_{\sigma} \varepsilon_i^{\rm PtPc}n_{i\sigma}^{\rm PtPc} \nonumber \\
	&&+\sum_{ij\sigma}[V_{ij}^{\rm Au-NaCl}(c^{\rm Au}_{i\sigma})^{\dagger}c^{\rm NaCl}_{j\sigma} +h.c.]  \\
	&&+\sum_{ij\sigma}[V_{ij}^{\rm NaCl-PtPc}(c^{\rm NaCl}_{i\sigma})^{\dagger}c^{\rm PtPc}_{j\sigma}+h.c.] \nonumber
\end{eqnarray}
The first three terms encode the energies of the Au states, the NaCl states, and the PtPc states, respectively.
The last two terms encode the coupling between Au and NaCl, and between NaCl and PtPc, respectively.
This Hamiltonian is used for describing the Au-NaCl-PtPc system inside a matching plane at $ z_0=1 $ \text{\AA} outside the nuclei of the PtPc molecule.
The calculations account for the Cl character of the states in the NaCl band gap due to the Cl character of the NaCl valence and conduction band \cite{clarkAugmentedPlane1968, deboerOriginConduction1999AmericanJournalofPhysics, olssonScanningTunneling2005, leonAnionicCharacter2022NatCommun}.

Beyond the matching plane, we introduce cylindrical coordinates with the radial coordinate, $\rho$, the azimuthal angle, $\phi$, and a coordinate perpendicular to the surface, $z$.
We follow Tersoff and Hamann\cite{tersoffTheoryApplication1983Phys.Rev.Lett., tersoffTheoryScanning1985Phys.Rev.B} and assume that the important orbital on the tip is an $s$ orbital.
The tip has a local radius of curvature $R$.
The center of the curvature is located a distance $z$ from the surface, which is also the center of the tip $s$ orbital.
The tip apex is then at a distance $z-R$.
In the following, we do not specify the value of $R$, and present our results as a function of $z$, corresponding to a tip distance of $z-R$.
Furthermore, we neglect changes of the potential in the NaCl-PtPc-vacuum system induced by the tip. 

Under these assumptions, the tip does not explicitly factor into the calculations.
Coulomb interactions are also not explicitly taken into account.
However, the HOMO and LUMO positions are adjusted to the measured values determined from scanning tunneling spectroscopy (STS).
As long as the important PtPc states are the neutral ground-state and states with one extra electron or one hole, Coulomb effects and image potential effects are implicitly accounted for by using level positions determined by STS.
This formalism, however, is incapable of describing exciton effects.
The TB solutions for substrate-barrier-molecule complex are matched continuously to the vacuum solution outlined below.

For $z>z_0$, we assume that the potential $V(\rho,\phi,z)$ takes the work function value $V_0=4.3$~eV \cite{liSpontaneousDoping2015Nanoscale} inside a cylinder radius $\rho_0=12$ \AA, and positive infinity outside.
Then, the potential $V(\rho,\phi,z)$ can be expressed as follows:
\begin{equation}\label{eq:4}
V(\rho,\phi,z)=
\begin{cases}
V_0,& \text{if}~\rho \le \rho_0~\text{and}~z\ge z_0 \\
\infty,& \text{if}~\rho>\rho_0~\text{and}~z\ge z_0
\end{cases}
\end{equation}
The energy zero is set at the Fermi energy of the substrate.
The cylinder radius (12 \AA) significantly exceeds the distance from the cylinder axis to the outermost H atoms (7.6 \AA) or the outermost C atoms (6.6 \AA).
Consequently, it is then a reasonable assumption that the true wave function of the tunneling electrons is localized within the cylinder.
We introduce the Schr\"odinger equation for an electron with the energy $\varepsilon (<V_0)$:
\begin{widetext}
\begin{eqnarray}\label{eq:5}
	\left[-\left(\dfrac{\partial^2}{\partial z^2}+\dfrac{1}{\rho}\dfrac{\partial}{\partial \rho}+\dfrac{\partial^2}{\partial\rho^2}\right)+ \dfrac{1}{\rho^2}\dfrac{\partial^2}{\partial \phi^2}+V(\rho,\phi,z)\right] \psi(\rho,\phi,z) = \varepsilon\psi(\rho,\phi,z)
\end{eqnarray}
\end{widetext}
Here we have expressed energies in Rydberg units (Ryd = 13.6 eV) and lengths are given in Bohr radii ($a_0=0.529$ \AA).
A solution in vacuum outside the PtPc molecule can be written as follows:
\begin{widetext}
\begin{eqnarray}\label{eq:6}
	\psi(\rho,\phi,z,\varepsilon)=\sum_{mi}\left[c_{mi}^{\rm (s)}{\sin}(m \phi) +c_{mi}^{\rm (c)} 
	{\cos}(m\phi)\right] J_m[k_{mi}\rho]e^{-\kappa_{mi}(\varepsilon) (z-z_0)}
\end{eqnarray}
\end{widetext}
where $m$ is a non-negative integer, and $J_m$ is an integer Bessel function.
The coefficients $k_{mi}$ are defined such that $J_m[k_{mi}\rho_0]=0$, ensuring that the wave function is zero for $\rho=\rho_0$.
To obtain the correct energy $\varepsilon$, we require
\begin{equation}\label{eq:7}
	[\kappa_{mi}(\varepsilon)]^2=[k_{mi}]^2-(\varepsilon-V_0).
\end{equation}
For $m\ge 1$, $\sin(m\phi)$ and for $m\ge 0$ $cos(m\phi)$ describe the azimuthal angle $\phi$ dependence, while the Bessel functions $J_m[k_{mi}\rho]$, $i=1, 2, \dots$ describe the radial behavior for a given $m$ value. 

The factor $\exp{\left[-2\kappa_{mi}(\varepsilon) z\right]}$ describes the exponential decay of the square of the wave function in the $z$-direction.
This factor introduces a strong energy dependence via the energy dependence of $\kappa_{mi}(\varepsilon)$ in Eq.~(\ref{eq:7}).
The leading contribution results from $\exp{\left[-2\kappa_{01}(\varepsilon)(z-z_0)\right]}$.
This factor varies over two orders of magnitude as the energy ranges from the bottom of the electronic transport gap, at $-1.3$~eV, to the top, at $1.7$ eV.
Additional relative variations among the different components are induced by $\exp\{-2[\kappa_{mi}(\varepsilon)-\kappa_{01}(\varepsilon)](z-z_0)\}$, as discussed in Table~\ref{table:1} below.
These considerations are rooted in the assumption that the potential reaches its vacuum value directly outside the molecule.
In this approach we have neglected the potential from the tip.
This potential is substantially higher in the study of the LUMO compared to that for the HOMO.
If this potential were included in the calculation, the pronounced enhancement of the LUMO versus the HOMO would be substantially smaller.
Furthermore, the potential is lowered outside the molecule, which we have neglected.
The image potential also contributes beyond what is implicitly included in the HOMO and LUMO positions.
While these factors should alleviate the strong energy dependence, they are neglected here.
These dependencies exist alongside the effects that occur during the propagation through the NaCl buffer and the PtPc molecule.

\section{Results and Discussion}\label{sec:results}
\subsection{STM measurements}  
\begin{figure}
	\centering
    \includegraphics[width=\linewidth]{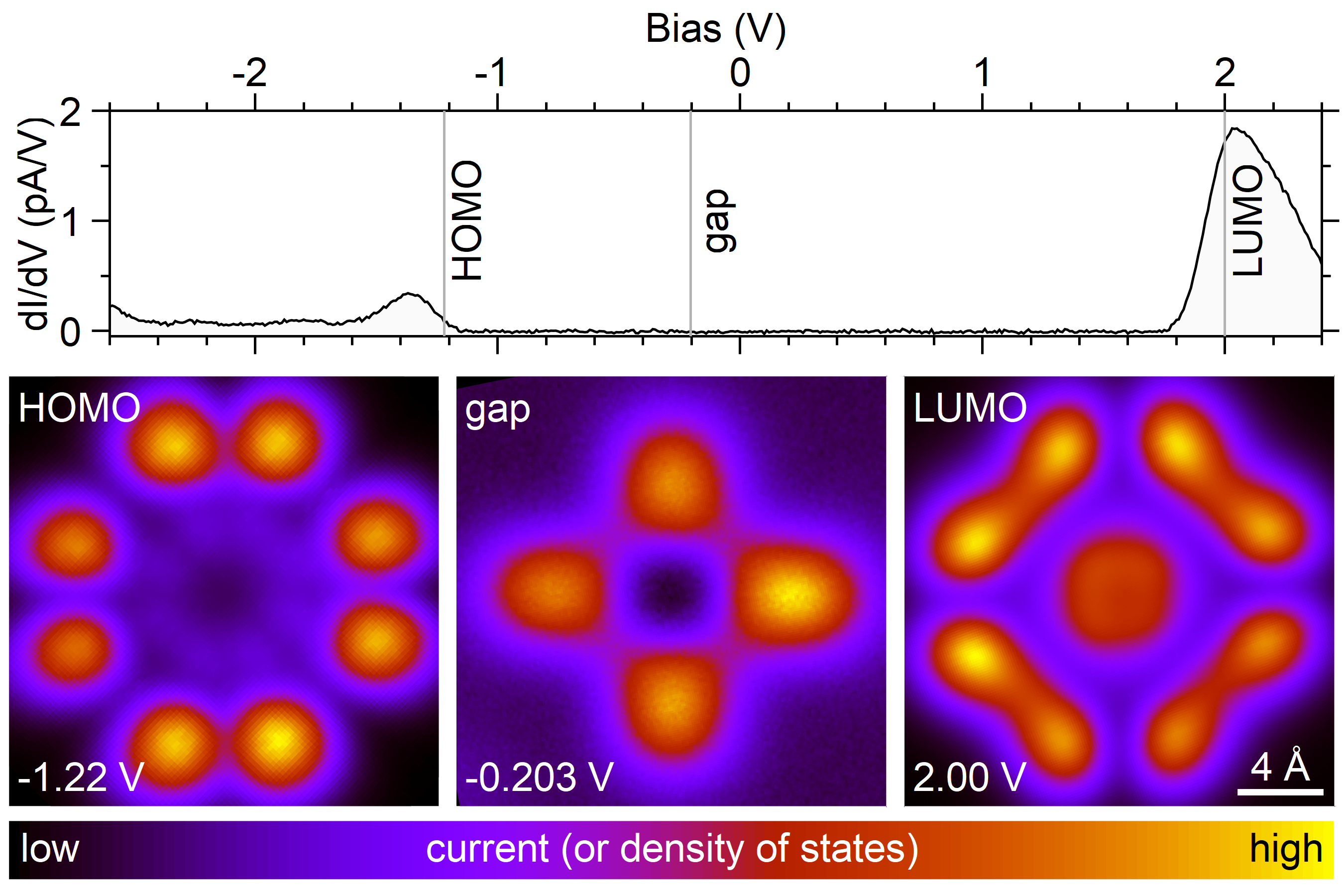}
	\caption{Typical differential conductance (d$I$/d$V$) spectrum on PtPc indicating the energies of the molecular frontier orbitals with broad local maxima at $-1.35$ V (HOMO) and $+2.05$ V (LUMO). Bottom row: Typical constant-height STM images of the frontier orbitals and of the feature appearing in the transport gap of the molecule. The color scale is valid for all figures with either measured (current) or calculated (DOS = density of states) images and spans the range from the minimum to the maximum of the absolute tunnel current. Here, the maxima of absolute currents are 66 pA (HOMO), 5.3 pA (gap feature), and 187 pA (LUMO). All images show an area of $ 20 \times 20~ \text{\AA}^2$.}
	\label{fig:2}
\end{figure}

The experiments were performed in a home-built low-temperature ($T\sim5$ K) STM operated under ultra-high vacuum at a base pressure of $\le 1\times10^{-11}$ mbar \cite{KuhnkeVersatileOptical2010RevSciInst}.
The preparation of single molecules of PtPc adsorbed on 3 monolayer NaCl(100) on Au(111) follows a procedure described previously \cite{grewalGapState2023ACSNano, GrewalSingleMolecule2022}.
All experimental data shown here are measured in constant-height mode.
The data in Fig.~\ref{fig:2} is presented for comparison with the theoretical calculations discussed below.
PtPc molecules are sensitive to the local inhomogeneity that results from the incommensurability of NaCl(100) with the herringbone reconstruction of Au(111).
Consequently, some molecules are more stably adsorbed and may thus be  preferentially selected by the experimentalist.
Additionally, the substrate causes some molecules to adsorb with the two-fold degeneracy of the LUMO lifted, enabling the imaging of one of the LUMOs in isolation.
We chose tip-sample distances that enable imaging molecules with a good signal-to-noise ratio while maintaining stable scanning conditions.
For excessively high currents and excessively short distances from the tip apex, molecules tend to move laterally, creating streaks in the image or even hop irreversibly to the tip.

The onset of this instability typically occurs at a tip-molecule distance of roughly 3~\AA, accompanied by a maximum tunnel current of 200 pA, with substantial variability in both values.
These values strongly depend on the applied voltage.
In the experiments presented here, the concern is not only to minimize the distance to the molecule but also to maximize the image contrast of the radial components of the molecular features.
The experimental images are intended to elucidate the effects of vacuum propagation in the electron tunneling process.

As we will demonstrate below, the constant height images (Fig.~\ref{fig:2}) obtained using the described method match very well with the calculations for molecule-tip distances ranging from 4-8~\AA. It is important to recall that we define the apex of the tip to be at $z-R$, where $R$ is curvature of the tip. Note also that the color bar used in Fig.~\ref{fig:2} is common to all 2D plots in this paper. Further details on its construction are given in SI.

\subsection{Theoretical results}
\begin{figure*}
	\centering
    \includegraphics[width = 340 pt]{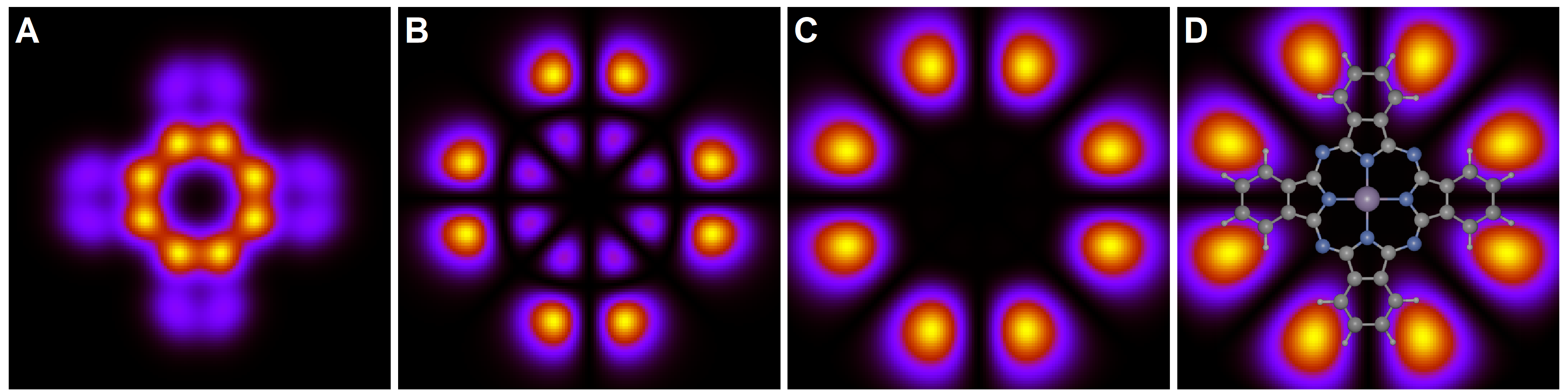}
	\caption{Calculated images at the energy of the HOMO ($\varepsilon=-1.3$ eV) for different distances from the molecular plane, $z-z_0$ = 0 $\text{\AA}$ (A), 4 $\text{\AA}$ (B), 8 $\text{\AA}$ (C) and 12 $\text{\AA}$ (D). All panels show an area of $ 20 \times 20~ \text{\AA}^2$. In panel D the molecular structure is superposed.}
	\label{fig:3}
\end{figure*}

We now present calculated images of a PtPc molecule.
Fig.~\ref{fig:3} shows the results for the HOMO ($\varepsilon=-1.3$~eV) for different values of $z-z_0$.
The molecule has four ``arms'' along the $x$- and $y$-axes, with the HOMO featuring nodes along these axes, resulting in eight lobe-like features as a function of azimuthal angle.
These lobes are particularly visible at larger $z$ values.
The HOMO image rapidly expands with increasing $z$ and undergoes noticeable changes in its shape. 
As discussed in the experimental section above, some $z$ values are too small to be experimentally accessible, yet observing the evolution with varying $z$ yields valuable insights. 

\begin{figure*}
    % ACS allows upto 504 pt wide figures. Set width option to \textwidth if necessary.
    \centering 
    {\includegraphics [width = 425 pt] {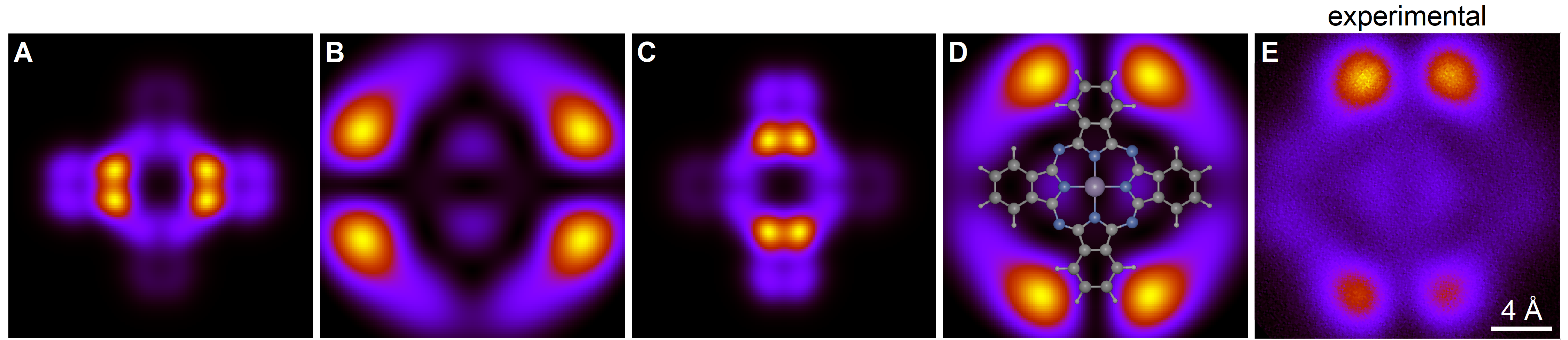}}
	\caption{Calculated images of the two individual LUMO degeneracy lifted orbitals ($\varepsilon=1.7$ eV) at the distances $z-z_0$ = 0 $\text{\AA}$ (panels A and C) and 12 $\text{\AA}$ (panels B and D). All panels show an area of $ 20 \times 20~ \text{\AA}^2  $. In panel D the molecular structure is superposed.
	Panel E shows a constant height STM image at a bias of only 1.35 V for a PtPc molecule for which the LUMO degeneracy happened to be lifted, probably due to a nearby substrate defect.}
	\label{fig:4}
\end{figure*}

\begin{figure*}
	\centering
    \includegraphics[width= 340 pt]{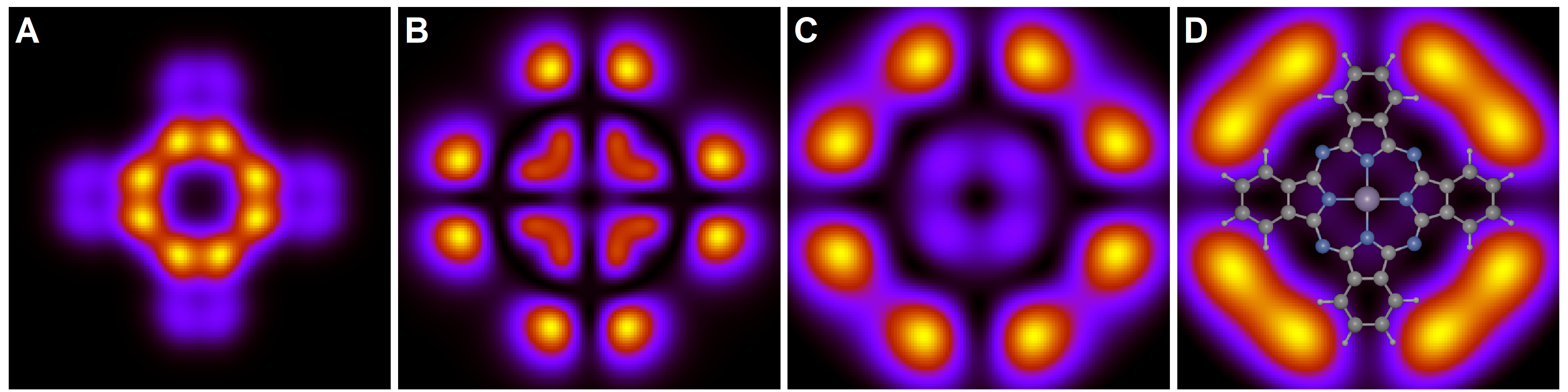}
	\caption{Calculated images of the sum of both LUMOs ($\varepsilon=1.7$ eV) for  distances, $z-z_0$ of 0 $\text{\AA}$ (A), 4 $\text{\AA}$ (B), 8 $\text{\AA}$ (C), and 12 $\text{\AA}$ (D). All panels show an area of $ 20\times20~ \text{\AA}^2  $. In panel D the molecular structure is superposed.}
	\label{fig:5}
\end{figure*}

Fig.~\ref{fig:4} display the results for the two degenerate LUMOs ($\varepsilon=1.7$ eV) separately, while Fig.~\ref{fig:5} presents their sum, corresponding to the typical topographical image obtained by STM.
One LUMO has most of its weight close to the $x$-axis, with a node along the $x$-axis.
The other LUMO has its weight distributed analogously along the $y$-axis.
The sum of the images results in eight features as a function of angle.
Interestingly, for $z-z_0=0$, the image bears a striking resemblance to the HOMO image at the same distance.
As $z$ increases, the LUMO image expands radially in a manner similar to the HOMO image.

The behavior of the angular features is quite different.
Note the behavior near the high symmetry axes.
For the HOMO, the eight angular features group into four pairs that straddle the positive and negative $x$- and $y$-axes.
As $z$ is increased, the features shift somewhat away from the $x$- and $y$ axis, with the maxima moving towards, e.g., $\pm 22.5^{\circ}$ and  minima at, e.g., $\pm 45^{\circ}$.
On the other hand, for the LUMO, this angular shift is much more important.
Substantial weight is built up at, e.g., $\pm 45^{\circ}$.
Strikingly, this implies that weight accumulates in directions where the underlying molecule has no atoms. 

\begin{figure*}
	\centering
    \includegraphics[width= 340 pt]{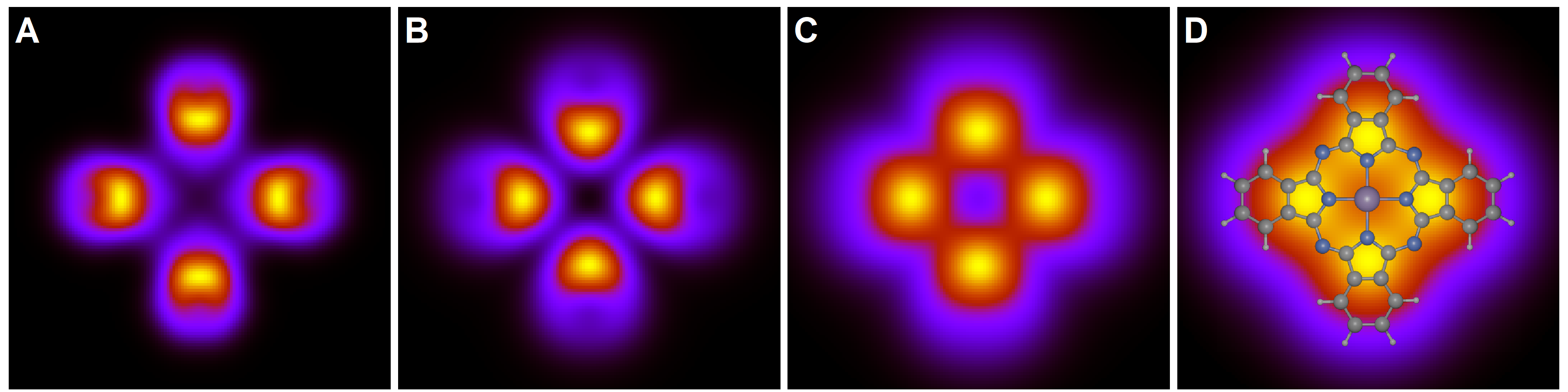}
	\caption{ Calculated images of the states at the substrate Fermi energy ($\varepsilon=0.0$ eV), that is in the molecular transport gap, for distances, $z-z_0$ of 0 $\text{\AA}$ (A), 4 $\text{\AA}$ (B), 8 $\text{\AA}$ (C), and 12 $\text{\AA}$ (D). All panels show an area of $ 20\times20~ \text{\AA}^2  $. In panel D the molecular structure is superposed.}
	\label{fig:6}
\end{figure*}

One might have expected that for energies between the HOMO and LUMO the image would show similarities to these MOs.
It is remarkable, however, that even for energies very close to either the HOMO or LUMO, the images look qualitatively distinct from their corresponding orbitals.
In contrast to the eight angular features of the HOMO and LUMO, only four features appear as a function of angle at voltages in the electronic transport gap.
The HOMO image has no weight, and the image comprising the two degenerate LUMOs has little weight along the $x$- and $y$-axes. 
In stark contrast, in-gap images have substantial weight built up along these axes.
Fig.~\ref{fig:6} shows that there is no clear radial expansion of the images with increasing $z$.
This behavior stands in stark contrast to the clear expansion observed for the HOMO and LUMO.

Similar results have been obtained in experiments for CuPc by Uhlmann \textit{et al.} \cite{uhlmannControllingOrbital2013NanoLett.} for the HOMO, LUMO, and the energies in between.
Our theoretical results for the HOMO and LUMO at $z-z_0=4~\text{\AA}~(z=5~\text{\AA})$ closely resemble the theoretical results obtained by Siegert \textit{et al.} \cite{siegertNonequilibriumSpin2016Phys.Rev.B} for CuPc at the same tip-sample distance.
However, theoretical results for energies lying between the HOMO and LUMO for CuPc, as reported in Ref.~\citenum{donariniTopographicalFingerprints2012Phys.Rev.B}, significantly deviate from both our theoretical and experimental results for PtPc.
These results also deviate from experimental findings by Uhlmann \textit{et al.} for CuPc \cite{uhlmannControllingOrbital2013NanoLett.}.
The reason is probably that the calculations in Ref.~\citenum{donariniTopographicalFingerprints2012Phys.Rev.B} did not include the lower-energy $\pi$-orbitals, which we demonstrate in the following to play a crucial role for topography images obtained at energies in the transport gap of the molecule.

\subsection{Detailed analysis of electron propagation effects}\label{sec:discussion}
We now turn our attention to the propagation of electrons through the vacuum region between the PtPc molecule and the tip.
This is done by examining the intricate details of the coefficients that describe the electron wave function.
In a forthcoming study, we will address the role of the propagation through the buffer (NaCl).

\begin{table}[b]
	\caption{$\kappa_{mi}(\varepsilon)$ (\AA$^{-1}$) determining the decay of the $z$-dependent 
	functions [Eq.~(\ref{eq:7})]. The numbers in parenthesis show the relative 
	damping exp$\lbrace -2[\kappa_{mi}(\varepsilon)-\kappa_{01}(\varepsilon)](z-z_0)\rbrace$ 
	of the intensity of a component relative to the $m=0$ and $i=1$ component for 
	$z-z_0=8$ \AA \ and $\varepsilon=0$.}
	\label{table:1}
	\begin{tabular}{rrrrrrr}
	\hline
	\hline
		$i$ &	$m=0$ & $m=1$ &  $m=2$ & $m=4$ & $m=8$  \\
	\hline
		1 & 1.08 (1.00) & 1.11 (.637) & 1.15 (.358) & 1.24 (.084) & 1.47 (.002) \\
		2 & 1.16 (.294) & 1.21 (.122) & 1.27 (.046) & 1.41 (.005) & 1.71 (.000) \\
		3 & 1.28 (.039) & 1.36 (.012) & 1.44 (.003) & 1.60 (.000) & 1.95 (.000) \\
		4 & 1.45 (.003) & 1.54 (.001) & 1.63 (.000) & 1.81 (.000) & 2.19 (.000) \\
	\hline
	\hline
	\end{tabular}
\end{table}

\subsubsection{Angular distortions and radial expansion}   
We first discuss the exponential factor $\exp{[-\kappa_{mi}(\varepsilon)z]}$, which describes the $z$-dependence in Eq.~(\ref{eq:6}) and Eq.~(\ref{eq:7}) of the wave function amplitude.
Table~\ref{table:1} presents the values of $\kappa_{mi}(\varepsilon)$. In parentheses we show the relative damping of the intensity of components $mi$     
\begin{equation}\label{eq:8}
	\exp{\{-2[\kappa_{mi}(\varepsilon)-\kappa_{01}(\varepsilon)](z-z_0)\}}
\end{equation}
relative to the $m=0$ and $i=1$ component for $(z-z_0)=8$~\AA \ and $\varepsilon=0$, close to the middle of the gap and well below the vacuum level at $V_0=4.3$~eV.
The table illustrates that i) components for small $m$ values are less damped and are thus strongly favored.
For each $m$ value, ii) components with small values of $i$ are strongly favored.
As discussed in the introduction, these effects emerge because a larger value of $m$ or $i$ leads to a larger kinetic energy in a plane parallel to the surface.
Since we study electrons of a given energy $\varepsilon$, the kinetic energy perpendicular to the surface is then correspondingly more negative and the decay in the $z$-direction exponentially more rapid.

We now turn to the Bessel functions describing the radial behavior that is the dependence on $\rho$.
Fig.~\ref{fig:7} shows these functions for $m$ = 0, 1 and 4.
With exception of small values of $m$, for e.g., $m=0$ and $m=1$, the function $J_m(k_{m1}\rho)$ primarily characterizes the outer regions of the molecule, encompassing the outermost C atoms and the surrounding space outside these atoms.
Conversely, the inner regions are predominantly characterized by $J_m(k_{mi})$ for $i>1$.

Below we demonstrate that the two effects previously discussed, i) and ii), along with the the corresponding behavior of the Bessel functions, are key for understanding the substantial distortion occuring in vacuum for the HOMO, LUMO and the in-gap images. 

\subsubsection{HOMO}
\begin{figure}
% {\rotatebox{0}{\resizebox{8.00cm}{!}{\includegraphics {plotweights/figureweightsz=9.eps}}}}
    \centering
    \includegraphics[width=200pt]{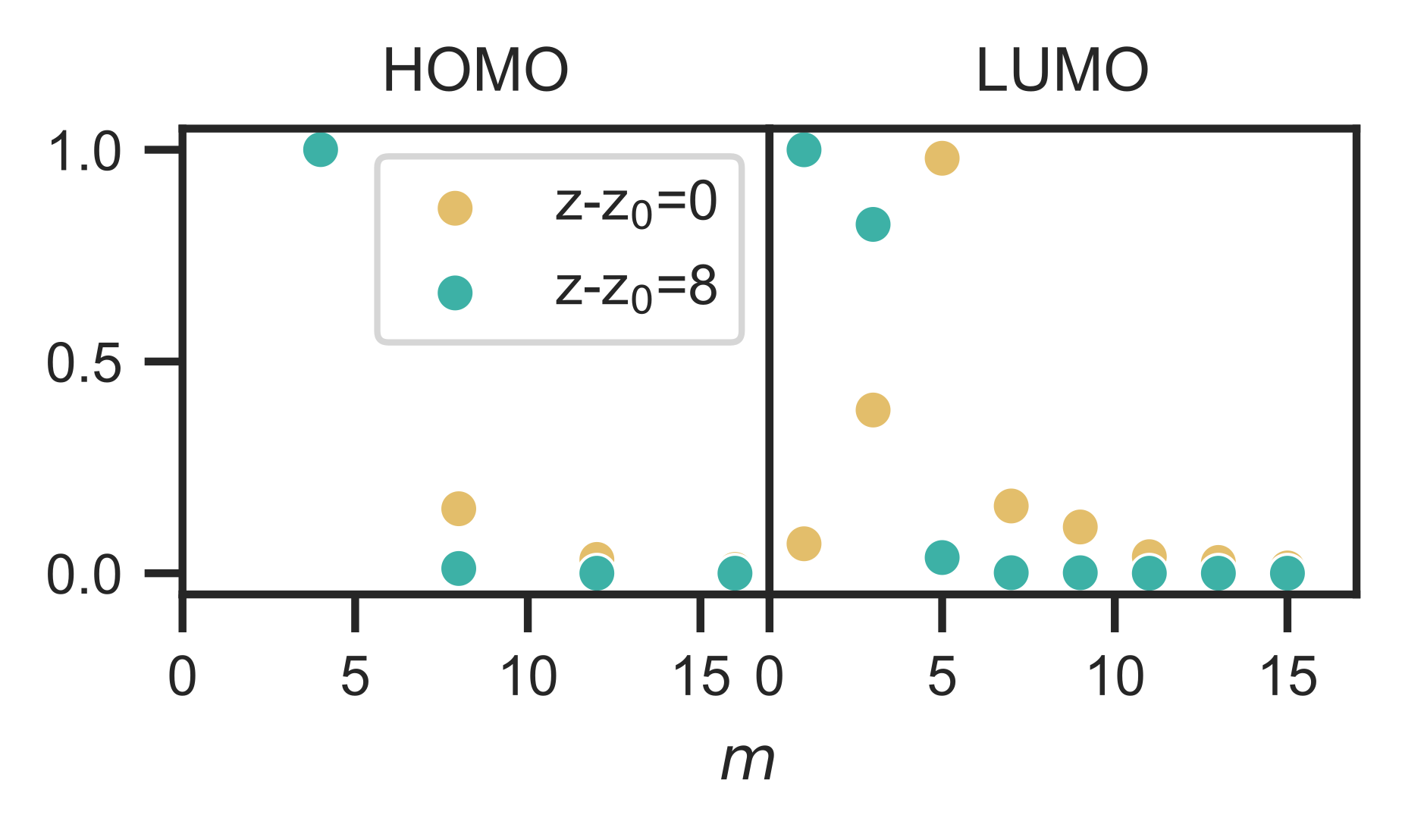}
	\caption{Relative weights of different $m$ components in images of the HOMO ($\varepsilon=-1.3$ eV) and LUMO ($\varepsilon=1.7$ eV) at distances $z-z_0=0$ and $8$ \AA. For each value of $z$, the results are normalized to the maximum value for that $z$. For the HOMO this is for $m=4$, for both values of $z$, while for the LUMO this is $m=5$ for $z-z_0=0$ and $m=1$ for $z-z_0=8$ \AA.}
	\label{fig:8}
\end{figure}

We first discuss the HOMO in more detail.
Given its eight features as a function of angle, it is  described by angular functions with $m=4\nu$, $\nu = 1, 2, \dots$.
The relative weight, 
\begin{equation}\label{eq:9}
S_m=\sum_i	[(c_{mi}^{\rm (s)})^2+(c_{mi}^{\rm (c)})^2]
	e^{-2[\kappa_{mi}(\varepsilon)-\kappa_{01}(\varepsilon)](z-z_0)},
\end{equation}
is shown in Fig.~\ref{fig:8} as a function of $m$ for $\varepsilon=\varepsilon_{\rm HOMO}$, highlighting the dominance of the $m=4$ terms over higher values of $m$ across all $z$ values. 

\begin{figure*}
	\centering
    \includegraphics[width=420 pt]{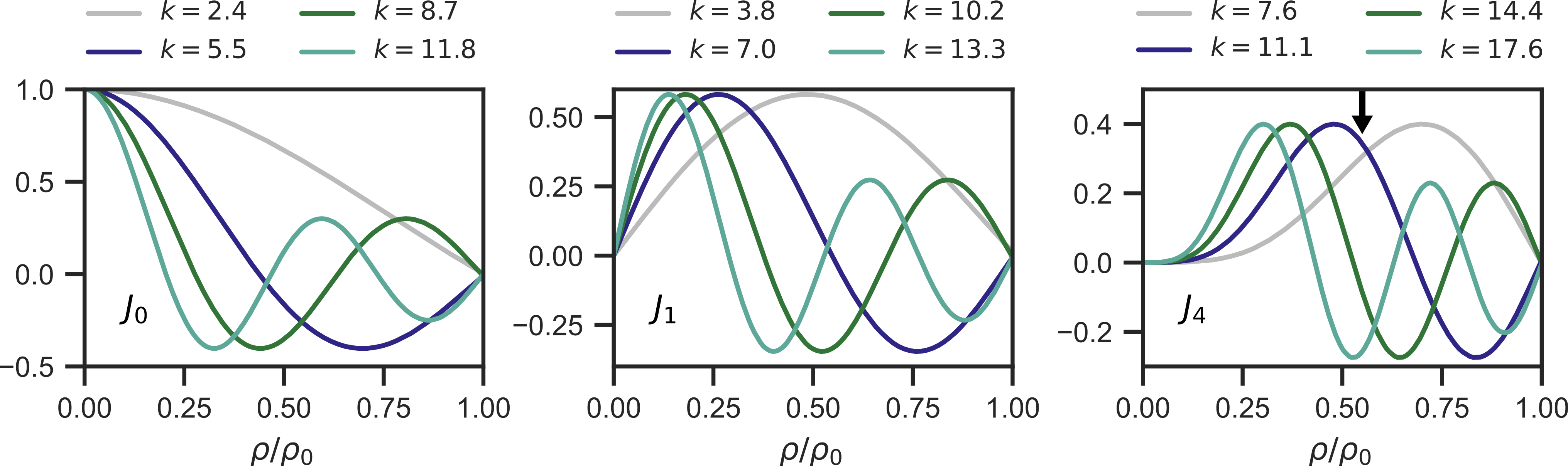}
    \caption{Bessel functions $J_m(k\rho/\rho_0)$ for $m$ = 0, 1 and 4 (from left to right) and for $k=k_{mi}$, $i$ = 1, 2, 3 and 4. The arrow in the right figure marks the positions of the outermost C atoms.}
    \label{fig:7}
\end{figure*}

We first focus on $m=4$.
The numbers in parenthesis in Table~\ref{table:1} show that the $i=1$ component is significantly less suppressed than the components with $i>1$ as $z$ increases.
This leads to a noticeable radial expansion of the image due to relatively higher weight of $J_4(k_{4i}\rho)$ for $i=1$ for large $\rho$ than the $i>1$ components (see Fig.~\ref{fig:7}).
Consequently, the centers of the eight HOMO lobes are positioned around $R=7.5$~\AA~($0.63~\rho_0$) for $z-z_0=8$~\AA, extending beyond the centers of the outermost C atoms at $R=6.6$~\AA~($0.55~\rho_0$)\footnote{These contributions are not due to the {{H}} atoms. {{Suppressing}} the contributions from {{H}} atoms made no visible change in the image. {{Actually}}, the {{MOs}} with {$\pi$} character do not couple to the {$1s$} {{H}} orbital, and {$\sigma$} orbitals, of importance in this context, have little {$1s$} {{H}} character.}.
The maximum of the image is at a somewhat smaller value of $\rho$ than the maximum ($\rho/\rho_0=0.70$) of $J_4(k_{41}\rho/\rho_0)$ due to the influence of mixed terms such as ones with $i=1$ and $2$, even for $z-z_0=8$~\AA.

The $m=4$ component results in eight intense features, prominently positioned at angles $\pm 22.5^\circ$ and separated by an angle of 45$^{\circ}$.
The higher order components, e.g., $m=8$, move these features pairwise together towards the $x$- and $y$-axes, as is illustrated for the case of $z=1$ \AA.
With increasing $z$, the influence of the $m>4$ components diminishes rapidly, causing the angular placement of the features in each quadrant to approach $\pm 22.5^\circ$.
However, even for $z-z_0=8$~\AA, these features are weakly displaced towards the $x$- and $y$-axes due to the non-negligible contributions from higher $m$-values.

\subsubsection{LUMO}
We now focus on the doubly degenerate LUMO.
Fig.~\ref{fig:5} presents the sum of the images of both states.
As mentioned earlier, HOMO and LUMO appear very similar for $z-z_0=0$, both undergoing significant expansion in the $\rho$-direction with increasing $z$.
However, their angular behavior diverge considerably.
Notably, the summation of the LUMOs builds up intensity along the angles $\pm45^{\circ}$ and $\pm135^{\circ}$ for large values of $z$.

To understand this build up, we first notice that the LUMOs are described by terms $\sin(m\phi)$ and $\cos(m\phi)$ = $\sin(m\phi + 90^{\circ})$ for odd values of $m$.
Let us consider the one LUMO with prominent features close to the $x$-axis, described by $\sin(m\phi)$ and shown in Fig.~\ref{fig:4} A, B.
The weights of its components with different $m$-values are shown in Fig.~\ref{fig:8}. 
For $z-z_0=0$, the $m=5$ component has the most weight.
This component has its maximum weight at the angle $90^{\circ}/5=18^{\circ}$.
As a result, this intensity appears approximately at $\pm 18^{\circ}$ for this LUMO, and naturally at $90^{\circ}\pm 18^{\circ}$ for the other LUMO (Fig.~\ref{fig:4}A, B).
The $m=3$ components are also relatively important, adding intensity that peaks at 90$^{\circ}/3=30^{\circ}$, thereby slightly shifting the overall intensity distribution away from the axes.
The two arms of one LUMO (at $\approx \pm 18^{\circ}$) are then about $36^{\circ}$ apart, while the two arms of two different LUMOs are about $90^{\circ}-36^{\circ}=54^{\circ}$ apart.
This leads to the features in Fig.~\ref{fig:5}, displaying both LUMOs, being relatively close to the $x$- and $y$-axes for $z-z_0=0$ \AA.

We can now turn our attention to $z-z_0=8$ \AA.
As demonstrated in Fig.~\ref{fig:8}, the $m=1$ and $m=3$ components are the largest.
For the LUMO with arms along the $x$-axis, these arms are shifted away from the axis, since $\sin(3\phi)$ has its maximum for a larger $\phi$ compared to, e.g., $\sin(5\phi)$, as evident in the right section of
Fig.~\ref{fig:4} (for $z-z_0=12$ \AA).
Similarly, for the other LUMO, the arms move away from the $y$-axis.
The result is the build-up of an appreciable weight around, e.g., $\phi=45^{\circ}$, even though there are no atoms in the underlying molecule in this direction.

\subsubsection{Gap states}

Finally, we consider states in the electronic transport gap.
These states were previously investigated in our earlier publication,\cite{grewalGapState2023ACSNano} wherein we concentrated on electron propagation from the substrate out to the PtPc molecule.
However, in the current work we focus on the propagation through the vacuum.
A striking result in Ref.~\citenum{grewalGapState2023ACSNano} was the rather simple image with four lobes along the $x$- and $y$-axes with no resemblance to the HOMO or LUMO, even though the measurement was performed at for energies relatively close to the HOMO or LUMO.
For the example of the state at the substrate Fermi energy (0.0 eV), Fig.~\ref{fig:6} illustrates that the weight of states within the transport gap of the molecule does not experience significant outward expansion. This is in contrast to the behavior observed in the case of the HOMO and the LUMO.

\begin{figure*}
\centering
\includegraphics[width=340 pt]{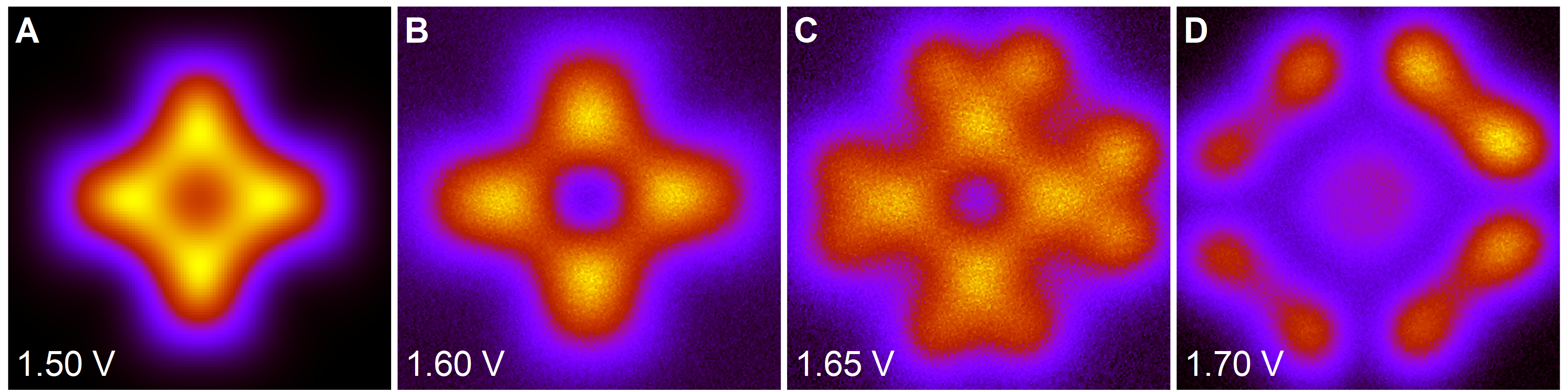}

	\caption{ Constant height STM images (B, C, D) illustrating the sharp transition between the gap feature and the LUMO. The bias voltages are indicated in each panel.	Panel A shows theoretical results for the bias 1.5 V. All panels show an area of $ 20\times20~\text{\AA}^2$.}
		\label{fig:SI1}
\end{figure*}
 
As discussed in Ref.~\citenum{grewalGapState2023ACSNano}, inside the PtPc molecule the states within the molecule's gap are to a good approximation linear combinations of bound PtPc states.
Very close in energy to the HOMO, the HOMO dominates the gap state, producing a HOMO-like image (not shown here).
However, moving up to energies above the HOMO, bound PtPc states below the HOMO and with small values of $m$ rapidly begin to contribute significantly to the image for $z=z_0$. 
The smallest $m$ value contributing to the HOMO is $m=4$, as shown in Table~\ref{table:1}, which efficiently undergoes exponential suppression as  $z$ is increased.
This suppression then causes states with $m=0$, 1 or 2 to gain relative prominence.
The exponential suppression effect remains influential even for energies a few tenths of an eV above the HOMO, rendering components with smaller $m$ values predominant in the image.
Despite the corresponding molecular states being considerably lower in energy than the HOMO, these smaller $m$ components dominate due to this strong suppression.
Analogous phenomena occur at energies below the LUMO. This is illustrated by the sharp and reversible transition from imaging the gap feature and the LUMO within a few tens of meV as shown in the STM measurements in Fig.\ref{fig:SI1} B, C, D. The theoretical result (panel A) corroborates that this transition indeed happens very rapidly as a function of tunneling energy.
Given the great importance of orbitals below the HOMO and LUMO for images within the transport gap, it is not surprising that these images could not be described in Ref.~\citenum{donariniTopographicalFingerprints2012Phys.Rev.B}, where these orbitals were not included.

We observe that the discussed effects render MOs below the HOMO very important for electron propagation at energies within the gap.
However, there is an additional physical mechanism at play.
Most of the MOs above the LUMO have many nodal surfaces, since they have to be orthogonal to lower states.
Consequently, these MOs then tend to have large kinetic energies parallel to the surface.
For a given bias voltage, they tend to have a very negative kinetic energy perpendicular to the surfaces, and therefore decay rapidly with increasing $z$ values.
This introduces a fundamental asymmetry between MOs well below versus most of those well above the gap.

For the HOMO and LUMO images, we found that the image strongly expands radially as $z-z_0$ increases from 0 to 12 \AA.
By comparison with Fig.\ref{fig:6} it becomes apparent that a similar strong expansion does not occur for the gap states.
For the HOMO, the expansion was attributed to the increasing dominance of the extended $i=1$ states as $z$ increases.
A similar predominance of small values of $i$ happens also for gap states.
However, for these states, the $m=0$ and $m=1$ components are particularly important, and within this range of $m$ values, the $i=1$ and the $i=2$ states have a comparable radial extension (see Fig.~\ref{fig:7})\footnote{If a much larger cylinder radius {$\rho_0$} is used, these arguments become more complicated, since the basis states for small values of {$i$} are then very extended. {{The}} numerical results for the images, however, are essentially unchanged.}.

\subsection{Vacuum propagation strongly favors certain MOs}
To gain a more comprehensive understanding of the origin of images for energies within the PtPc transport gap, we continue the PtPc MOs at specific energies $\varepsilon$ into the vacuum region.
The continuation, $\Psi_{\nu}(\rho,\phi,z,\varepsilon)$, for MO $\nu$, is formulated in terms of the basis functions employed in Eq.~(\ref{eq:6}).
The function $\Psi_{\nu}$ matches the MO $\nu$ continuously at $z=z_0$.
Next, we consider a solution $\tilde \Psi_{\alpha}$ of the combined system comprising the substrate-buffer-PtPc and the vacuum at an energy $\varepsilon_{\alpha}$. 
Within the vacuum we write the solution as 
\begin{equation}\label{eq:2}
	\tilde \Psi_\alpha(\rho,\phi,z)=\sum_{\nu=1}^{182}a^{(\alpha)}_{\nu}
	\Psi_{\nu}(\rho,\phi,z,\varepsilon_{\alpha})
\end{equation}
We then calculate the overlap integral, $S_{\nu,\mu}(z)$ of two functions $\Psi_{\nu}$ and $\Psi_{\mu}$, over a plane parallel to the surface and at a distance $z$ from the molecule.
Finally, we introduce a density matrix
\begin{equation}\label{eq:3}
	P_{\nu,\mu}=C\sum_{-0.8\le \varepsilon_{\alpha}\le 1.2}[a^{\alpha}_{\nu}]^{\ast}
	a^{\alpha}_{\mu}S_{\nu,\mu}
\end{equation}
where we sum over all states within the energy range from $-0.8~eV$ to $1.2~eV$ in the transport gap.
To ensure normalization, the constant $C$ is chosen so that 
\begin{equation}\label{eq:10}
	\sum_{\nu,\mu=1}^{182}P_{\nu\mu}(z)=1.
\end{equation}
It is notable that some of the off-diagonal terms might be negative.
This matrix illustrates how the propagation through the different MOs contributes to the image. 

\begin{figure}[b]
    \centering
    \includegraphics[width=166.66pt]{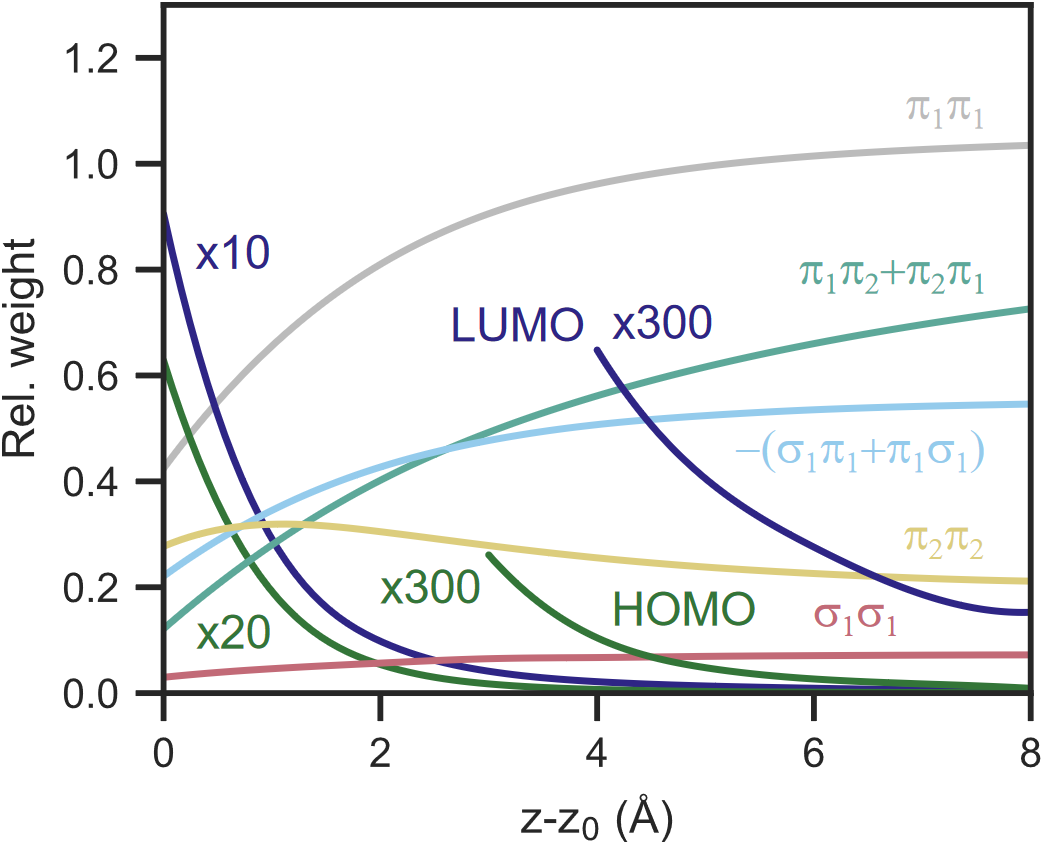}
	\caption{Theory predicted weights $P_{\nu,\mu}(z)$ describing contributions to the image due to products of tails of important MOs. The product was integrated over a plane at $z - z_0$ and over the bias range (-0.8 eV $\le \varepsilon \le 1.2$ eV). We show diagonal elements, $P_{\nu\nu}$, for the lowest $\pi$-orbital ($\pi_1$, $\varepsilon-\varepsilon_F=-8$ eV), a $\pi$-orbital with an approximately cylindrical nodal surface ($\pi_2$, $\varepsilon-\varepsilon_F=-5$ eV), the diagonal terms summed over 7 orbitals about 1.8 eV below the HOMO as well as the HOMO and LUMO. The latter are plotted twice, each with the indicated enhancement factor. We also show two off-diagonal contributions involving the lowest $\sigma$ orbital ($\sigma_1$, $\varepsilon-\varepsilon_F=-26$ eV) and $\pi_1$-orbital as well as the $\pi_1$ and $\pi_2$ orbitals.	}
	\label{fig:9}
\end{figure}

\begin{figure*}
	\centering
    \includegraphics[width=340 pt]{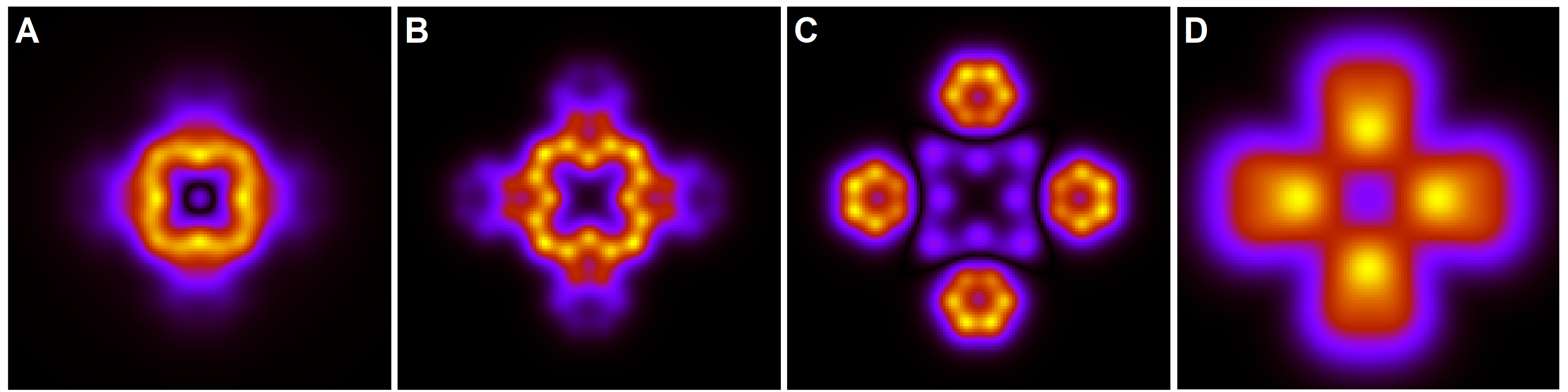}
	\caption{Important MOs for images in the gap. The panels show the absolute value of (A) the lowest $\sigma$ MO ($\sigma_1$ in Fig.~\ref{fig:9}), (B) the lowest $\pi$ MO ($\pi_1$), and (C) a $\pi$ MO with one ``radial'' node 
	($\pi_2$). The second MO ($\pi_1$) is positive everywhere (for $z>0$), while the first MO ($\sigma_1$) is slightly negative at parts very close to the Pt nucleus ($\le 0.7 \text{\AA}$) due to the Pt $3z^2-r^2$ d-orbital. We emphasize the similarity of the $\sigma_1$ and $\pi_1$ orbitals. The inner parts of $\pi_2$ are negative and the outer parts are positive. In constructing the images from Fig.~\ref{fig:3} to Fig.~\ref{fig:6} the coefficient of the $\sigma$ orbital has the opposite sign to the coefficients of the two $\pi$-orbitals which leads to an overall cancellation of weight in the inner parts of the image. The orbitals were calculated at $z-z_0=0~\text{\AA}$. 
	Panel D shows the calculated image at -1.0 V (a value just above the HOMO energy at -1.3 V) for $z-z_0=6$ \AA.
All panels show an area of $ 20\times20~\text{\AA}^2$.}
	\label{fig:10}
\end{figure*}

Fig.~\ref{fig:9} shows the $z$-dependence of important elements of $P_{\nu\mu}$.
Some of the corresponding orbitals are shown in Fig.~\ref{fig:10}. 
Remarkably, even the lowest ($\sigma_1$) state, at -26 eV, contributes more significantly than the HOMO (-1.3 eV) and the LUMO (1.7 eV).
As previously discussed, this phenomenon stems from to the lowest $\sigma$ orbital having only a small positive kinetic energy in the plane of the molecule, due to the absence of nodes in this plane.
For a given total energy $\varepsilon$, the kinetic energy perpendicular to the molecule is then not very negative, and the exponential decay in vacuum not so rapid.  

Similar effects happen for the lowest $\pi$ orbital, $\pi_1$, although its smaller energy difference to $\varepsilon_F=0$ results in a significantly larger contribution.
The $\pi_2$ orbital has a radial node (as seen in Fig.~\ref{fig:10}), resulting in a somewhat larger kinetic energy parallel to the surface, a more negative kinetic energy perpendicular to the surface, and consequently, a more rapid decay in the vacuum. Panel D shows the calculated image at a value just above the HOMO energy. It indeed shows great similarities with a linear combination of the orbitals in panels B and C.

The HOMO has components with $m=4$ or higher $m$ values.
Table~\ref{table:1} shows that the $m=4$ components decay rapidly.
As shown in the SI, most of the HOMO weight is furthermore in components with fairly large values of $i\sim 4$.
This configuration results in the kinetic energy parallel to the surface being very large, and a corresponding rapid decay outside the molecule. 
We have also summed the diagonal contributions from 7 orbitals (``7 orbital''), within a calculated energy range of approximately $0.1$ eV, and around $1.8$ eV below the HOMO.
These orbitals have much more weight than the HOMO and the LUMO, rendering them potentially important in some specific contexts.
The SI shows that the leading components of the LUMO are also characterized by large values of $m$ and $i$, and the corresponding rapid decay in vacuum.
However, the LUMO also has components with small values of $m$ and $i$.
Despite these components' small amplitudes at the matching surface ($z-z_0=1$ \AA), their relatively slow decay in vacuum results in their significant contribution.
Furthermore, much of the contribution from the LUMO comes from the upper part of the studied energy interval, where the wave functions generally decay more slowly.
Consequently, the LUMO contribution ($4\times10^{-4}$ from both LUMOs) is not as drastically reduced during the propagation through vacuum as observed forthe HOMO contribution ($2\times10^{-5}$).
The observation that the most important contributions stem from orbitals lacking angular nodes (as shown in Fig.~\ref{fig:9}) elucidates the shapes of the images presented in Fig.~\ref{fig:6}.

\section{Summary}\label{sec:summary}
We have analysed the transport of electrons through the vacuum region in STM studies of PtPc adsorbed on a thin NaCl film atop an Au substrate.
For propagation through vacuum, basis functions with few nodes in the angular and radial directions parallel to the surface are favored.
Such functions exhibit small kinetic energy in these directions, subsequently leading to a less negative kinetic energy perpendicular to the surface at a given bias voltage.
This results in their slower exponential decay in the vacuum region.
These effects are directly influenced by the thickness of the vacuum layer, i.e., the distance to the tip.
When considering the HOMO or LUMO, both experimental and theoretical analyses reveal a substantial radial expansion of the image.
This expansion is attributable to the dominance of vacuum basis functions that lack radial nodes.
In the case of the LUMO, there is also a substantial angular distortion, due to the additional emphasis of basis functions with few angular nodes.
This leads to a substantial weight along the diagonals $x=y$ and $x=-y$, despite the underlying molecule having no atoms at these positions.
The analysis moreover demonstrates that the low resolution observed for orbitals with many nodes has its physical origin in the electron propagation into vacuum and is not exclusively imposed by the limited spatial resolution of the STM tip.
For propagation within the transport gap, the image is due to a linear combination of waves originating from many MOs.
Notably, the significance of HOMO and LUMO quickly reduces when entering the transport gap, already at energies just a few tenths of an eV above the HOMO or below the LUMO energy, leading to a radical change in the resulting STM topography image.
Waves characterized by few angular and radial nodes then make a large contribution.
Particularly, this accentuates the role of $\pi$-MOs located energetically well below the HOMO.
These intricate details provide a consistent and rational basis for understanding the investigative capabilities of the STM.

%\newpage

%\begin{suppinfo}
%	Tight-binding model - NaCl(100) on Au(111), Tight-binding model - PtPc, Propagation in vacuum,
%	Colour bar generation.
%\end{suppinfo}

%\section{Data availability}
%The data supporting this study's findings are available from the corresponding authors upon reasonable request.

%\section{Competing interests statement}
%The authors declare no competing interests.

%\section{Author contributions statement}
%All authors participated in the analysis and discussion of the results, and in the writing of the manuscript.
%The tight-binding calculations were planned and carried out by O.G..
%A.G. and C.C.L. performed the experiments.
%O.G., K.Ku., and K.Ke. conceived and designed the research program.

\bibliography{manuscript}
\end{document}

%\date{}
\title{Supplementary Information for\\ Scanning Tunnelling Microscopy for Molecules: Effects of Electron\\ Propagation into Vacuum}

\maketitle

\subsection*{Tight-binding model -- NaCl(100) on Au(111)}
We use the same model for NaCl on Au as in our earlier work \cite{grewalGapState2023ACSNano}.
This model consists of a NaCl(100) three-layer film on a Au(111) substrate.
The NaCl film contains $9\times 9 \times 4$ atoms per layer.
We use three different clusters representing the Au substrate and average the results.
The three Au clusters have four, six, or eight layers with 1020, 780, and 572 Au atoms per layer, respectively.
In total there are then 486 Na, 486 Cl, and 4080, 4680, or 4576 Au atoms.
We impose periodic boundary conditions parallel to the surface for both, the NaCl slab and the Au slabs.
Hopping integrals are constructed according to the nearest neighbour hopping rules rules of Harrison \cite{harrisonElementaryElectronic1999, harrison1980electronic}, including $s-d$ \cite{harrison1980electronic} and $p-d$ hopping.
The NaCl and Au surfaces are non commensurate.
We place the central Na atom on top of a central Au atom in the neighbouring NaCl and Au layers.

To describe the Au substrate, we use the lattice parameter $a_{\rm Au}=4.07$ \AA~\cite{daveyPrecisionMeasurements1925Phys.Rev.}.
We use the Harrison level energies $\varepsilon_{6s}=-6.98$ eV and $\varepsilon_{5d}=-17.78$ eV as a reference and add a $6p$ level at 5 eV above the $4s$-level.
We then shift the $5d$ level relative to the $6s$ and $6p$ levels so that the top of the $5d$ band is placed at 1.7 eV below the Fermi energy \cite{sheverdyaevaEnergymomentumMapping2016Phys.Rev.B}.
Finally all energies are shifted so that the Fermi energy is at zero.
The resulting parameters are summarised in Table~\ref{table:1}.

To describe the NaCl film, we essentially follow our earlier work \cite{leonAnionicCharacter2022NatCommun, grewalGapState2023ACSNano} and choose parameters such that the conduction band has mainly Cl $4s$ character \cite{leonAnionicCharacter2022NatCommun, deboerOriginConduction1999AmericanJournalofPhysics}.
To achieve this, we replace the Cl $3s$ level by a $4s$ level, which has been strongly lowered by the Madelung potential, while the Na levels are strongly shifted upwards.
We fine-tune the Harrison parameters to replicate the experimentally observed bulk NaCl band gap (8.5 eV)
\cite{pooleElectronicBand1975Phys.Rev.B}.
Image potential effects were neglected.
According to calculations \cite{wangQuantumDots2017NanoLett.} using the GW method \cite{hedinNewMethod1965Phys.Rev.}, the top of the valence band is 5 eV below the Fermi energy.
We then shift all the NaCl energies relative to the Au energies correspondingly.
The resulting parameters are shown in Table~\ref{table:1}.
The calculations were performed for lattice parameter $a_{\rm NaCl}=5.54$ \AA~\cite{chenPropertiesTwodimensional2014Phys.Chem.Chem.Phys.}.

We use the calculated separation $d_{\rm Au-NaCl}=3.12$ \AA~between the Au surface and the NaCl film \cite{chenPropertiesTwodimensional2014Phys.Chem.Chem.Phys.}.
Since the NaCl film and the substrate are incommensurate, several Au atoms can have similar distances to a given NaCl atom, and the nearest neighbours are poorly defined.
We then use a smooth distance dependent cut-off of the Harrison prescription for the hopping between the substrate and the film.
Thus, the Harrison prescription for these hopping integrals is multiplied by a factor
\begin{equation}\label{eq:s7}
\exp{\left(\frac{-(d-d_{\rm Au-NaCl})^2}{\lambda_{\rm SB}^2}\right)},
\end{equation}
where $d$ is the distance between an Au atom and a NaCl atom at the Au-NaCl interface.
Here $\lambda_{\rm SB}$ is chosen such that summing these factors over all the Au neighbours of a NaCl atom and averaged over the NaCl atoms in the innermost layer adds up to four.
Then the innermost NaCl atoms effectively couple to four Au atoms.

\begin{table}[t]
\begin{tabular}{lccc}
    \hline
    \hline
    Element & $s$ & $p$ & $d$ \\
    \hline
    Au (6s, 6p, 5d) & 4.1 & 9.1 & $-3.7$ \\
    Na (3s, 3p) & 12.8 & 16.8 & $-$ \\
    Cl (4s, 3p) & 10.2 & $-5.0$ & $-$ \\
    C (2s, 2p) & $-19.38$ & $-11.07$ & $-$ \\
    N (2s, 2p) & $-26.22$ & $-13.84$ & $-$ \\
    Pt (6s, 5d) & $-6.85$ & $-$ & $-16.47$ \\
    H (1s) & $-13.61$ & $-$ & $-$ \\
    \hline
    \hline
\end{tabular}
\caption{Level energies used for PtPc, NaCl, and Au after shifts described in the text.
The Fermi energy is put at zero.}
\label{table:1}
\end{table}

\subsection*{Tight-binding model -- PtPc}
We study the absorbed molecule platinum phthalocyanine (PtPc).
The coordinates are obtained from a density functional calculation.
The tight-binding parameters are primarily obtained from Harrison \cite{harrisonElementaryElectronic1999}.
For the H atoms we include the $1s$ level at the energy $-13.6$ eV (not given by Harrison).
Guided by Miwa {\it et al.} we use the separation 3.4 \AA~between the the PtPc molecule and the NaCl film.
PtPc is absorbed atop a Na atom \cite{miwaEffectsMoleculeinsulator2016Phys.Rev.B}.
For PtPc the four ``arms'' of the molecule are along the NaCl (100) directions.
For the hopping between the molecule and NaCl, we also use the rules of Harrison \cite{harrisonElementaryElectronic1999}, but modified as in Eq.~(\ref{eq:s7}) above.
Again a $\lambda_{BM}$ is chosen so that, on average, from each atom in the molecule there is effectively hopping to four atoms in the NaCl buffer.
The Au slab breaks the four-fold symmetry of PtPc which has been reintroduced in the plots.

These parameters incorrectly puts a $\sigma$-orbital below the HOMO.
We therefore shift this orbital by 3.2 eV upwards, slightly above the LUMO.
We also adjust the parameters so that the experimental gap is obtained, including image effects.
Finally we align the levels with the Fermi energy ($E_F=0$) of the system, so that the HOMO is located at $-1.3$ eV and LUMO at 1.7 eV, in agreement with experiment.
The resulting parameters are shown in Table \ref{table:1}.

\begin{figure*}[t]
\includegraphics[width=270pt]{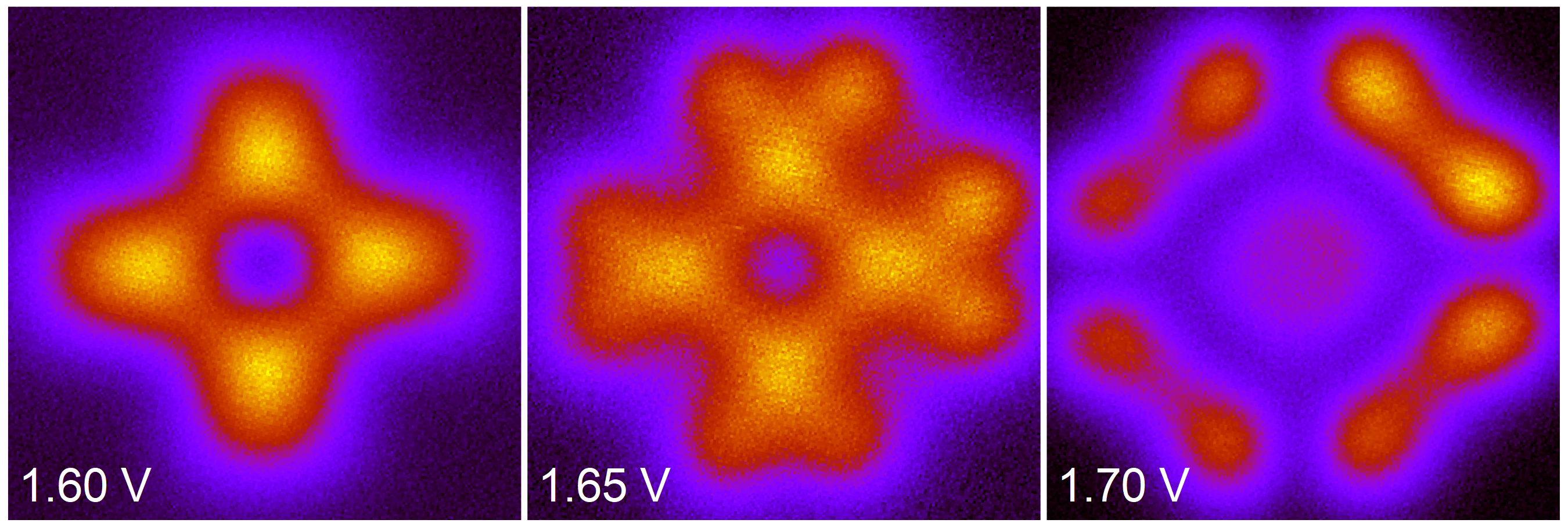}
\caption{Constant height STM images illustrating the sharp transition between the gap feature and the LUMO. The bias voltages are indicated in each panel. All panels show an area of $ 20\times20~\text{\AA}^2$.}
\label{fig:SI1}
\end{figure*}

\subsection*{Propagation in vacuum}
In Eq.~(3) in the main text, we presented the expansion of the MOs of PtPc in basis function appropriate for vacuum propagation
\begin{equation}\label{eq:1}
	\sum_{mi}\left[c_{mi}^{\rm (s)}{\rm sin}\left(m \phi\right) +c_{mi}^{\rm (c)}
	{\rm cos}(m\phi)\right]
	J_m\left[k_{mi}\rho\right]e^{-\kappa_{mi} z},
\end{equation}
where $m(\ge 0)$ is an integer and $J_m$ is an integer Bessel function.
The values of $\kappa_{mi}$, determining the exponential decay with $z$, were given in Table I in the main text.
As discussed in that context, the different components $mi$ decay at very different rates in vacuum, with the $m=0$ and $i=1$ component decaying most slowly.
Table~\ref{table:2} shows the expansion coefficients for some important MOs in terms of these vacuum functions.
The largest component for the lowest $\sigma$ and lowest $\pi$ orbitals is the $m=0$ $i=1$ component, which, therefore, decay relatively slowly with $z$.
The $\pi$ orbital with one radial node has somewhat larger amplitudes for higher components, and decays somewhat faster.
The HOMO only has components with $m=4$ or higher $m$ values, which decay rapidly with $z$.
In particular, most of these components, in addition, have rather high values of $i$, making the decay even faster.
The LUMO has the largest amplitude for an $m=5$ component with a high $i$ value.
This component decays very rapidly.
However, there are also components with $m=1$ and $m=3$, which have small amplitudes, but are still important because of their slower decay.
Overall, the LUMO therefore does not decay as rapidly as the HOMO.

\begin{table}
\begin{tabular}{ccccccccccccccc}
    \hline
    \hline
    \multicolumn{3}{c}{{\rm Lowest} $\sigma$ } & \multicolumn{3}{c}{{\rm Lowest} $\pi$} & \multicolumn{3}{c}{{\rm Higher} $\pi$ } & \multicolumn{3}{c}{\rm HOMO} & \multicolumn{3}{c}{\rm LUMO} \\
    $m$ & $i$ & $c_{mi}^{\rm (c)}$ & $m$ & $i$ & $c_{mi}^{\rm (c)}$ & $m$ & $i$ & $c_{mi}^{(c)}$ & $m$ & $i$ & $c_{mi}^{\rm (s)}$ & $m$ & $i$ & $c_{mi}^{(\rm c)}$ \vspace{0.25em}\\
    \hline
    0 & 1 & 1.00 & 0 & 1 & 1.00 & 0 & 1 & 0.81 & 4 & 1 & $-0.23$ & 1 & 1 & $-0.03$ \\
    0 & 2 & 0.78 & 0 & 2 & 0.73 & 0 & 3 & $-0.92$ & 4 & 3 & 0.41 & 1 & 2 & 0.13 \\
    0 & 4 & $-0.32$ & 0 & 4 & $-0.36$ & 0 & 5 & 0.47 & 4 & 4 & 1.00 & 3 & 4 & $-0.69$ \\
    0 & 5 & $-0.46$ & 0 & 5 & $-0.40$ & 4 & 1 & 0.62 & 4 & 5 & 0.90 & 5 & 4 & 1.00 \\
    0 & 6 & $-0.36$ & 0 & 6 & $-0.25$ & 4 & 2 & 1.00 & 4 & 6 & 0.42 & 5 & 5 & 0.97 \\
    0 & 7 & $-0.16$ & 4 & 5 & 0.17 & 4 & 3 & 0.47 & 8 & 3 & $-0.38$ & 5 & 6 & 0.56 \\
    \hline
    \hline
\end{tabular}
\caption{Largest expansion coefficients for a few MOs [Eq.~(\ref{eq:1})].
In addition, we show a few coefficients for small $i$ and $m$ values.
The table shows results for the lowest $\sigma$ MO, the lowest $\pi$ MO, a $\pi$ MO with one radial node as well as the HOMO and one LUMO.
For a given orbital, the coefficients have been renormalized so that the largest coefficient is unity.}
\label{table:2}
\end{table}

\subsection*{Transition from gap image to LUMO}
As an example of how abrupt the transition from the orbital image to the transport gap feature occurs, we show in Fig.~\ref{fig:SI1} images that were recorded with voltage differences of only 50 mV. We remark that a similarly sharp transition was observed also for the transition from HOMO to the gap feature at negative voltages.

\subsection*{Colour bar generation}
Let $\gamma = 1.25$ and the range of $x$ be $[0, 1]$. The colour bars of the main text are parameterised by the following equations for each primary colour component. 
\begin{eqnarray*}
	\textrm{red}(x) &&= x^{\frac{\gamma}{2}} \\
	\textrm{green}(x) &&= x^{3\gamma} \\
	\textrm{blue}(x) &&= \left\{
	\begin{array}{ll}
	\sin(2\pi x^\gamma), & 0\leq x^\gamma \leq 0.5 \\
      0, & 0.5 < x^\gamma \leq 1\\
	\end{array}
	\right.
\end{eqnarray*}

\bibliography{si}
\vfill